\newcounter{myctr}
\def\myitem{\refstepcounter{myctr}\bibfont\noindent\ifnum\themyctr>9\else\phantom{0}\fi\hangindent17pt\themyctr.\enskip}

\documentclass{ws-ijqi}
\usepackage{hyperref}
\usepackage[super,sort,compress]{cite}

\usepackage{bbm}
\usepackage{color}

\newcommand{\upe}{\mathrm{e}}

\newcommand{\SNR}{\mathrm{SNR}}
\newcommand{\FQd}{F_\mathrm{Q}(d)}
\newcommand{\nb}{n_\mathrm{b}}
\newcommand{\ns}{n_\mathrm{s}}
\newcommand{\sinc}{\mathrm{sinc}}
\newcommand{\dhalf}{d_{1/2}}
\newcommand{\kbar}{\bar{k}}

\begin{document}

\catchline{}{}{}{}{}

\title{Resolution limits of spatial mode demultiplexing with noisy detection}

\author{Yink Loong Len and Chandan Datta}

\address{Centre for Quantum Optical Technologies, Centre of New Technologies, University of Warsaw,\\
Banacha 2c, 02-097, Warsaw, Poland\\
y.len@cent.uw.edu.pl; c.datta@cent.uw.edu.pl}

\author{Micha\l{} Parniak and Konrad Banaszek}

\address{Centre for Quantum Optical Technologies, Centre of New Technologies, University of Warsaw,\\
Banacha 2c, 02-097, Warsaw, Poland}
\address{Faculty of Physics, University of Warsaw,\\
Pasteura 5, 02-093 Warszawa, Poland\\
m.parniak@cent.uw.edu.pl; k.banaszek@cent.uw.edu.pl}

\maketitle

\begin{history}
\received{\today}
\end{history}

\begin{abstract}

We consider the problem of estimating the spatial separation between two mutually incoherent point light sources using the super-resolution imaging technique based on spatial mode demultiplexing with noisy detectors. We show that in the presence of noise the resolution of the measurement is limited by the signal-to-noise ratio (SNR) and the minimum resolvable spatial separation has a characteristic dependence of $\sim\SNR^{-1/2}$. Several detection techniques, including direct photon counting, as well as homodyne and heterodyne detection are considered.
\end{abstract}

\keywords{Optical imaging; quantum estimation; super-resolution}


\section{Introduction}\label{Sec1}	
Imaging is a routine task in scientific activities, ranging from detecting fluorescence for molecular structures \cite{RalphPittetNATS2008,RagVDBNM2008}, to observing and resolving stellar images.\cite{LabeyrieAA1970} The standard method of direct imaging, i.e., registering intensity distribution in the image plane with the help of optical instruments suffers resolution limits due to diffraction from finite apertures of the optical instruments. Various clever strategies and solutions have been introduced and implemented to improve the resolution beyond the diffraction limits; see Refs.~\citen{LabeyrieAA1970,RamWardPNAS2006,DonnertKellerPNAS2006,GiovannettiLyodPRA2009,TsangPRL2009,ThielBastinPRA2009,PicheLeachOE2012,XuSongAPL2015,KellererRibakOL2016,TsangNairPRX2016,NairTsangPRL2016,PaurStoklasaOPTICA2016} for a selected representation of these works. In particular, in Ref.~\refcite{TsangNairPRX2016}, Tsang \emph{et. al.} have introduced the {technique of spatial mode demultiplexing (SPADE), which is able to provide resolution that is beyond the diffraction limits and approaches the ultimate limit derived in accordance to the quantum theory. In the case of imaging two point sources, the super-resolving power in the proposal of Tsang \emph{et. al.} is most evident, if we include certain assumptions, in particular that the point sources have equal brightness and are mutually incoherent, the centroid is known, and the detectors are perfect. The consequences and importance of some of these assumptions, and the possibility of relaxing or incorporating them, have been a subject of a number of recent works.\cite{ChrostowskiDDIJQI2017,RehacekHradilPRA2017,RehacekHradilPRA2018,LarsonSalehOPTICA2018,ParniakBorowkaPRL2018,TsangNairOPTICA2019,LarsonSalehOPTICA2019,FisherSteinbergNJP2019,GraceDutton2019,HradilRehacek2019}
\bigskip

In this paper we address the effects of detection noise on the super-resolving power of the SPADE measurement. The organization of this paper are as follows. In Sec.~\ref{Sec2}, we set up the model and introduce the problem. We will consider determination of the distance between two point sources, which serves as an elementary model to discuss resolution limits of imaging. Then, we introduce the generic tools that will be used to analyze the problem, via the example of direct imaging method. In Sec.~\ref{Sec3}, we review the SPADE measurement as suggested by Tsang \emph{et. al.}, and demonstrate how it achieves the promised superresolution, when the  detection process is noise-free. We present the main results of this work in Sec.~\ref{Sec4} and Sec.~\ref{Sec5}, where the effect of noise on measurement with SPADE is studied respectively for photon-counting and quadrature measurements. We show that in the presence of noise the superresolution offered by the SPADE measurement is lost for sources that are too close, and as a rule of thumb, the minimum resolvable spatial separation has a dependence of $\sim\SNR^{-1/2}$, where $\SNR$ stands for signal-to-noise ratio. Finally, Sec.~\ref{Sec6} concludes the paper.

\begin{figure}
\centering
\includegraphics[width=0.65\columnwidth]{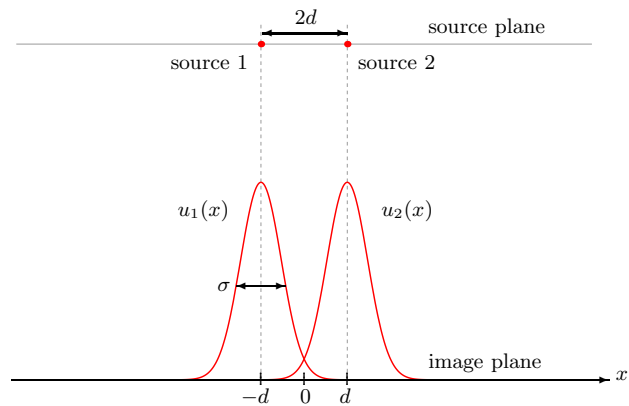}
\caption{Two point sources forming image in one spatial dimension, where the image plane is parallel to the source plane. Due to limited resolution of the imaging apparatus, each source generates a non-singular coherent field distribution described by the transfer function $u_j(x)$, which has a characteristic width $\sigma$,  at the image plane. For simplicity, 1:1 magnification of the imaging system has been assumed in this figure.}
\label{Fig1}
\end{figure}

\section{Preliminaries}\label{Sec2}
Consider two mutually incoherent point sources, labelled with index $j=1,2$ respectively. To focus on the effects of noise, we will make the familiar assumptions that \emph{a priori} the source intensities are equal and the centroid is known. Moreover, for simplicity we will discuss
image formation in one spatial dimension, where our measurement will be along the image plane which is parallel to the source plane, as depicted in Fig.~\ref{Fig1}. Due to the diffraction limits from the objectives, each source generates a non-singular coherent
field distribution described by an amplitude transfer function $u_j(x)=u(x-x_j)$, with its peak located at $x_1=-d$ or $x_2=d$. The transfer function satisfies $\int_{-\infty}^\infty\,\mathrm{d}x\, |u_j(x)|^2=\int_{-\infty}^\infty\,\mathrm{d}x\, |u(x)|^2=1$, and $|u_j(x)|^2$ identifies the probability density of a given photon emitted by source $j$ being detected at position $x$ at the image plane. To simplify calculations, we will assume from now on that $u(x)$ is real. Our interest is to determine the spatial separation between the sources, particularly in the ``small separation regime'', i.e., when $d$ is much smaller than the characteristic width of the transfer function, $\sigma$ (to be defined precisely later). Eventually, the knowledge of $d$ is to be translated to the knowledge of the separation in the source plane, where the one-to-one mapping between them, typically a function of the effective focal length of the instrument objectives and the distance from the sources to the screen, are assumed to have been well calibrated and known. In the temporal domain, we shall be mainly concerned with a situation where the image is essentially built up from a series of repeated and statistically independent single-photon detection events over some total period much larger than the coherence time of the sources. We will also consider a scenario where only single temporal mode of the electromagnetic field is detected and the effective source statistics is thermal. \medskip

With the specifications above, the commonly employed direct imaging method, i.e., registering the light intensities at each pixels (taken to be infinitely dense) at the image plane, turns out to be a very inefficient way of extracting the information about $d$ in the small separation regime, and hence resolving the two sources, in the sense that the estimator for $d$ will be subjected to large uncertainty. Formally, for any unbiased estimator, such as the popular maximum-likelihood estimator in the asymptotic limit of large data\cite{kay1993fundamentals}, the precision, quantified by the root-mean-square error $\Delta d$, can at most be reduced to as small as $[F(d)]^{-1/2}$, where $F(d)$ is the Fisher Information (FI) for $d$,
\begin{align}
F(d)=\ns\int_{-\infty}^{\infty}\,\mathrm{d}x\, \frac{1}{p(x)} \Big(\frac{\partial}{\partial d} p(x)\Big)^2, \label{eq1}
\end{align}
where
\begin{align}
p(x)=\frac{1}{2}|u_1(x)|^2+\frac{1}{2}|u_2(x)|^2=\frac{1}{2}|u(x-d)|^2+\frac{1}{2}|u(x+d)|^2 \label{eq2}
\end{align}
is the probability density of a given photon detected at $x$, and $\ns$ is the mean total photon number emitted by the sources over the total observation period $T$. For $d$ sufficiently small such that we can approximate $u_{1,2}(x)=u(x\pm d)\approx u(x)\pm d u'(x)+\frac{d^2}{2}u''(x)$, we have
\begin{align}
F(d)\approx 4\ns d^2 \int_{-\infty}^{\infty}\,\mathrm{d}x\, \Big(\frac{[u'(x)]^2}{u(x)}+u''(x)\Big)^2
\propto 
\ns d^2,  \label{eq3}
\end{align}
and therefore $\Delta d\gtrsim \ns^{-1/2}d^{-1}$, which increases unlimitedly as $d\rightarrow0$. This can be understood intuitively, since direct imaging method can hardly distinguish between two closely overlapping transfer functions and one corresponding to a single point source located at the center of the two sources. On the other hand, in the large separation regime where the two transfer functions have virtually zero overlap, the FI for direct imaging is
\begin{align}
F(d)
&\approx \ns \int_{-\infty}^{0}\,\mathrm{d}x\, \frac{1}{\frac{1}{2}[u_1(x)]^2}  \Big(\frac{\partial}{\partial d} \frac{1}{2}[u_1(x)]^2\Big)^2 + \ns \int_{0}^{\infty}\,\mathrm{d}x\, \frac{1}{\frac{1}{2}[u_2(x)]^2}  \Big(\frac{\partial}{\partial d} \frac{1}{2}[u_2(x)]^2\Big)^2\nonumber\\
&\approx 2\ns \int_{-\infty}^{0}\,\mathrm{d}x\, [u_1'(x)]^2 + 2\ns \int_{0}^{\infty}\,\mathrm{d}x\, [u_2'(x)]^2 \approx 4\ns \int_{-\infty}^{\infty}\,\mathrm{d}x\, [u'(x)]^2,  \label{eq4}
\end{align}
which is finite.

\noindent For benchmarking, we shall make reference to the Quantum Fisher Information (QFI), $\FQd$, which is the highest FI obtainable by optimizing over all physical measurement strategies.\cite{helstrom1976quantum} It can be shown that for independently emitted individual photons,\cite{TsangNairPRX2016}
\begin{align}
\FQd = 4\ns \int_{-\infty}^{\infty}\,\mathrm{d}x\, [u'(x)]^2 \equiv \frac{\ns}{\sigma^2},  \label{eq5}
\end{align}
where
\begin{align}
\sigma\equiv \frac{1}{2}\Big(\int_{-\infty}^\infty\,\mathrm{d}x\, [u'(x)]^2\Big)^{-1/2}.  \label{eq6}
\end{align}
Comparing Eq.~(\ref{eq5}) to Eq.~(\ref{eq4}), we see that when the two transfer functions are well separated, direct imaging is an excellent choice, and there is little room for improvement by considering other measurement strategies. We will therefore be interested in this work in the small separation regime, where direct imaging is proven to be ineffective. In addition, Eq.~(\ref{eq5}), with the implication $\Delta(\frac{d}{\sigma})\geq\ns^{-1/2}$, allows us to identify $\sigma$ defined in Eq.~(\ref{eq6}) as the intrinsic physical scale for the problem, and thus can be taken as the natural definition of the characteristic width of the transfer function.

\section{SPADE with photon counting measurement}\label{Sec3}
The measurement with spatial mode demultiplexing (SPADE), introduced in Ref.~\citen{TsangNairPRX2016}, has been shown to achieve near-optimal precision in the small separation regime. The basic idea of super-resolution imaging based on SPADE is to measure the intensity of incoming radiation in a basis of normalized spatial modes $\{v_i(x)\}_{i=0,1,\dots}$, chosen suitably according to the transfer function. One mode is the transfer function itself, $v_0(x)\equiv u(x)$, and another one is proportional to the
derivative of the transfer function,
\begin{align}
v_1(x)\equiv -2\sigma \frac{\mathrm{d}u}{\mathrm{d}x}=-2\sigma u'(x), \label{eq7}
\end{align}
where $\sigma$ is given by Eq.~(\ref{eq6}). One can verify readily that $\{v_0(x),v_1(x)\}$ are orthonormal function set, i.e., $\int_{-\infty}^\infty\,\mathrm{d}x\, v_i(x)v_j(x)=\delta_{i,j}$, where, as a reminder, real transfer functions are assumed here. By applying for example Gram-Schmidt procedure to higher order derivatives of the transfer function, one can thus complete an orthonormal spatial mode basis. Experimentally, the decomposition into modes $\{v_i(x)\}$ can be achieved by using integrated optics waveguide structures,\cite{TsangNairPRX2016} a spatial light modulator, \cite{PaurStoklasaOPTICA2016} or a multi-plane light converter \cite{MorizurNicholls2010,LabroilleDenolle2014}.

By the assumptions made in Sec.~\ref{Sec2}, photons emitted by the sources are randomly and independently sorted into the modes $v_j(x)$, with the probability or transmission coefficient
\begin{align}
\tau_j(d)&=\frac{1}{2}\Big|\int_{-\infty}^\infty\,\mathrm{d}x\, v_j(x)u_1(x)\Big|^2+\frac{1}{2}\Big|\int_{-\infty}^\infty\,\mathrm{d}x\, v_j(x)u_2(x)\Big|^2, \label{eq8}
\end{align}
and subsequently detected by photon counting. Since these detection events are mutually independent, the total FI is additive, i.e., $F(d)=\sum_{j=0}^\infty F_j(d)$, where $F_j(d)\geq0$ is the FI contributed by detection from mode $v_j(x)$.

\subsection{Small separation regime: Binary SPADE}
For the small separation regime where $d/\sigma\ll1$, which is of our main concern, it suffices to consider binary SPADE with just two modes, $v_0(x)$ and $v_1(x)$, as the photons detected in the $v_1(x)$ mode would carry most information about $d$. To see this, first observe that over the whole detection period $T$, we have on average $\kbar(d)\equiv\ns\tau_1(d)$ photons measured in the mode $v_1(x)$.
Next, the transmission coefficient for the mode $v_1(x)$ can be approximated by
\begin{align}
\tau_1(d)&\approx\frac{1}{2}\Big|\int_{-\infty}^\infty\,\mathrm{d}x\, v_1(x)[u(x)-du'(x)]\Big|^2+\frac{1}{2}\Big|\int_{-\infty}^\infty\,\mathrm{d}x\, v_1(x)[u(x)+du'(x)]\Big|^2 \nonumber\\
&=\frac{d^2}{4\sigma^2}, \label{eq9}
\end{align}
where $\approx$ stands for equality up to the leading order in series expansion in $d/\sigma$. Then, if the photon emission statistics from the sources is Poissonian with mean $\ns$, using Eq.~(\ref{eqA2}) in Appendix, the FI is
\begin{align}
F_1^{\textsc{(p)}}(d)=\ns\frac{1}{\tau_1}\Big(\frac{\partial}{\partial d}\tau_1\Big)^2\approx\frac{\ns}{\sigma^2}=\FQd. \label{eq10}
\end{align}
Alternatively, if the light detected from the source exhibits Bose-Einstein statistics, we have, by Eq.~(\ref{eqA4}) in Appendix,
\begin{align}
F_1^{\textsc{(t)}}(d)=\frac{1}{1+\kbar}F_1^{\textsc{(p)}}(d)\approx \Big(\frac{1}{1+\frac{\ns d^2}{4\sigma^2}}\Big)\frac{\ns}{\sigma^2}\approx \FQd, \label{eq11}
\end{align}
if we have sufficiently small $d$, i.e., $d/\sigma\ll1$, and $d/\sigma\ll2/\sqrt{\ns}$ such that $\kbar\ll1$. \medskip

\noindent More generally, regardless of the transfer function and the photon emission statistics, as long as $\kbar(d)\ll1$, which is always the case for finite $\ns$ and small enough $d$, two or more photon detection events will almost be unobserved and hence negligible. Then, it is sufficient to consider just the single-photon detection events which happen with probability $p_1\approx\kbar$, and we have
\begin{align}
F_1(d)\approx \frac{1}{p_1}\Big(\frac{\partial}{\partial d}p_1\Big)^2\approx \frac{\ns}{\sigma^2}=\FQd, && [\ns\tau_1\ll1]\label{eq12}
\end{align}
confirming that binary SPADE as the adequate measurement to use for small enough $d$.
In the limit of ever smaller $d$, $F_1(d)\rightarrow\frac{\ns}{\sigma^2}=\FQd$ remains finite, and therefore one is able to estimate arbitrarily small $d$ with finite and near-optimal precisions---the gist and essence of super-resolution feature of SPADE. \medskip

\begin{figure}[t!]
	\centering
	\includegraphics[width=1\columnwidth]{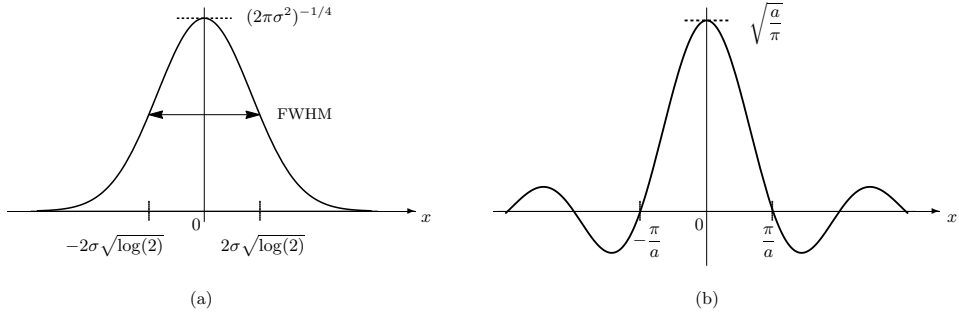}
	\caption{(a) Gaussian transfer function $u(x)=(2\pi \sigma^2)^{-1/4}\upe^{-\frac{x^2}{4\sigma^2}}$, characterized by its full-width-at-half-maximum (FWHM) which is proportional to $\sigma$. (b) Sinc transfer function $\sqrt{a/\pi}\,\sinc(ax)$, characterized by $\sigma=\sqrt{3}/(2a)$, where $2\pi/a$ measures the size of the main lobe of the function.}
	\label{Fig2}
\end{figure}

 We close this section by a brief illustration with the examples of Gaussian and sinc transfer functions, both with two kinds of photon emission statistics, namely the Poissonian and thermal or Bose-Einstein distribution. For the Gaussian transfer function,commonly used as an example of a regularized transfer function, we have
\begin{align}
u(x)&=\frac{1}{(2\pi \sigma^2)^{1/4}}\upe^{-\frac{x^2}{4\sigma^2}}, \hspace{0.5cm} v_1(x)=\frac{x}{\sigma}u(x), \label{eq13}
\end{align}
where $\sigma$, consistent with the definition in Eq.~(\ref{eq6}), has here the usual meaning of the standard deviation of the probability density $[u(x)]^2$, or equivalently, $4\sigma\sqrt{\log(2)}$ measures the full-width-at-half-maximum of the transfer function, as shown in Fig.~\ref{Fig2}. For the sinc transfer function, which is produced by diffraction from a hard aperture in one dimension, we have
\begin{align}
u(x)&=\sqrt{\frac{a}{\pi}}\,\sinc(ax)=\sqrt{\frac{a}{\pi}}\,\frac{\sin(ax)}{ax}, \hspace{0.5cm} v_1(x)=2\sigma\sqrt{\frac{a}{\pi}}\frac{1}{x}\Big(\sinc(ax)-\cos(ax)\Big),  \label{eq14}
\end{align}
where $a=\frac{\sqrt{3}}{2\sigma}$. As depicted in Fig.~\ref{Fig2}, the significance of $a$, and hence of $\sigma$, is that $\pm \frac{\pi}{a}$ characterizes the width of the main lobe of the sinc transfer function, and it is related to the physical parameters by $a=2\pi\delta y/\lambda$, where $\lambda$ is the wave length of the light, $2\delta$ is the slit width, and $y$ is the image conjugate distance.

\begin{figure}[t]
	\centering
	\includegraphics[width=1\columnwidth]{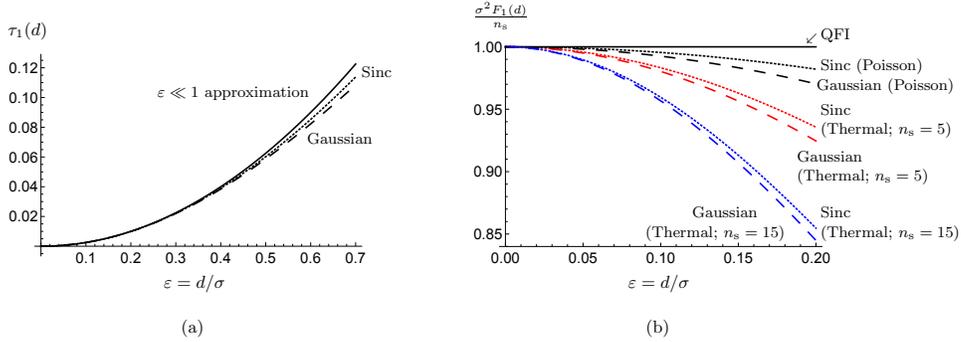}
	\caption{(a) The transmission coefficients for the mode $v_1(x)$ with Gaussian transfer function (dashed line), and sinc transfer function (dotted line). The  quadratic approximation for small separation regime, Eq.~(\ref{eq9}), is plotted as the solid line. (b) The Fisher Information per mean photon number, multiplied by $\sigma^2$, for Gaussian (dashed) and sinc transfer functions (dotted), and with a Poissonian distribution (black) and thermal distributions with $\ns=5$ (red) and $15$ (blue), respectively. As can be seen, while the values of the Fisher Information are different for different transfer functions and photon emission statistics, they always converge to the Quantum Fisher Information (QFI) for sufficiently small $d$.}
	\label{Fig3}
\end{figure}

The transmission coefficient for mode $v_1(x)$, given by Eq.~(\ref{eq8}), and as plotted in Fig.~\ref{Fig3} for the two transfer functions, are respectively given by
\begin{align}
\tau_1(d)&=\frac{d^2}{4\sigma^2}\upe^{-\frac{d^2}{4\sigma^2}}
\approx\frac{d^2}{4\sigma^2}\Big(1-\frac{d^2}{4\sigma^2}\Big), && \text{[Gaussian]}  \label{eq15} \\
\tau_1(d)&=\frac{16d^4}{3\sigma^4}\Big( \sin(ad)-ad\cos(ad) \Big)^2
\approx\frac{d^2}{4\sigma^2}\Big(1-\frac{3d^2}{20\sigma^2}\Big). && \text{[Sinc]}  \label{eq16}
\end{align}
Then, using Eq.~(\ref{eq15}), Eq.~(\ref{eq16}) and Eq.~(\ref{eq10}),  we have
\begin{align}
F_1^{\textsc{(p)}}(d)&=\ns\frac{1}{\sigma^2}\upe^{-\frac{d^2}{4\sigma^2}} \Big(1-\frac{d^2}{4\sigma^2}\Big)^2
\approx \frac{\ns}{\sigma^2} \Big(1-\frac{3d^2}{4\sigma^2}\Big), && \text{[Gaussian]}  \label{eq17} \\
F_1^{\textsc{(p)}}(d)&=\ns\frac{4\sigma^2}{d^4}\Big( 4\cos(ad)-4\sinc(ad)+2ad\sin(ad) \Big)^2 \nonumber\\
&\approx\frac{\ns}{\sigma^2} \Big(1-\frac{9d^2}{20\sigma^2}\Big), && \text{[Sinc]}  \label{eq18}
\end{align}
and finally $F_1^{\textsc{(t)}}(d)=\frac{1}{1+\ns\tau_1(d)}F_1^{\textsc{(p)}}(d)$ by Eq.~(\ref{eq11}).  The plots of these transmission coefficients and the corresponding FIs are shown in Fig.~\ref{Fig3}.

\section{Binary SPADE with noisy photon-counting measurement}\label{Sec4}
As mentioned in the Introduction, in obtaining Eqs.~(\ref{eq10}-\ref{eq12}) that signify the super-resolution feature of binary SPADE, we have implicitly made the assumption that there is no noise at all throughout the whole detection period. This is, however, at best an approximation to the realistic situation, as total exclusion and elimination of noise is impossible. Intuitively, one expects that noisy detection will degrade the performance of SPADE measurement, and set a limit to the resolution and precision that we can achieve in the laboratory. In this section, we will provide a characterization and quantification to the limit of resolution of a noisy SPADE measurement with photon counting.
\medskip

Specifically, we consider the most typical, yet crucial noise that one encounters in photon-counting experiments, namely the background or dark count noise, i.e., unwanted photoelectric events registered by the detectors which are not originated from the light sources in investigation. A common source of the background counts is the random thermal excitations of the photoelectrons in the detector itself, which produce photoelectric current independent of the presence of the light sources. In addition, the detectors might pick up stray photons from the environment which adds to the total number of detection events.

Denote $\nb$ as the mean photocount number contributed by the noise to the detection in mode $v_1(x)$ over the detection time $T$, and define the signal-to-noise ratio (SNR) as the ratio between the mean photon from the sources, $\ns$, to $\nb$: $\SNR\equiv \ns/\nb$. With $\beta\equiv {1}/{\SNR}$, the total mean photon number registered in mode $v_1(x)$ in time $T$ is therefore
\begin{align}
\kbar(d)=\ns\tau_1(d)+\nb=\ns\big[\tau_1(d)+\beta\big],\label{eq19}
\end{align}	
and when the resultant statistics is either Poisson or thermal distribution, the FI for mode $v_1(x)$ is respectively
\begin{align}
F_1^{\textsc{(p)}}(d)&=\frac{1}{\kbar}\Big(\frac{\partial}{\partial d}\kbar\Big)^2=\frac{\ns}{\tau_1+\beta}\Big(\frac{\partial}{\partial d}\tau_1\Big)^2,  \label{eq20} \\
F_1^{\textsc{(t)}}(d)&=\frac{1}{\kbar+1} F^{\textsc{(p)}}(d)=\frac{\ns}{(1+\ns\tau_1+\nb)(\tau_1+\beta)}\Big(\frac{\partial}{\partial d}\tau_1\Big)^2.   \label{eq21}
\end{align}
Using Eq.~(\ref{eq9}) for the small separation regime, we have
\begin{align}
F_1^{\textsc{(p)}}(d)&\approx \frac{\ns}{\sigma^2}\Big( \frac{1}{1+\frac{4\sigma^2}{d^2}\beta} \Big), \label{eq22}\\
F_1^{\textsc{(t)}}(d)&\approx \frac{\ns}{\sigma^2}\Big( \frac{1}{1+\frac{4\sigma^2}{d^2}\beta} \Big) \Big( \frac{1}{1+\frac{\ns d^2}{4\sigma^2}+\nb}\Big). \label{eq23}
\end{align}

\begin{figure}[t!]
	\centering
	\includegraphics[width=0.9\columnwidth]{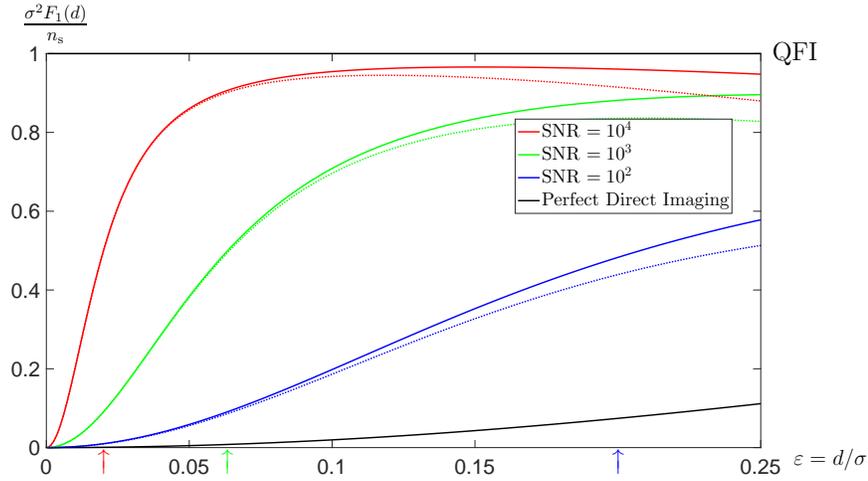}
	\caption{Binary SPADE with noisy photon-counting detection: FI per mean incoming photon $\ns$, multiplied by $\sigma^2$, for different signal-to-noise ratios (SNR), with Gaussian transfer function. Solid curves are with Poissonian sources, while dotted curves are with thermal sources. For the thermal source case, $\ns=5$ was assumed. As shown, superresolution for arbitrarily small $d$ is now lost, and we characterize the resolution limits by $\dhalf\equiv\frac{2\sigma}{\sqrt{\SNR}}$, which is the smallest separation that can be estimated with at least $\frac{1}{\sqrt{2}}$ of the optimal precision. In the figure, the limits are marked by the three arrows with the corresponding color codes. For comparison, we have included the FI computed for noiseless direct imaging (the black solid curve).}
	\label{Fig4}
\end{figure}

Comparing Eq.~(\ref{eq22}) and Eq.~(\ref{eq23}) to Eq.~(\ref{eq10}) and Eq.~(\ref{eq11}), the presence of noise introduces an additional factor $\big(1+\frac{4\sigma^2}{d^2}\beta\big)^{-1}$, and it qualitatively changes the  behaviour of $F_1(d)$. In particular, for sufficiently small $d$, this additional factor scales as $d^2$, which in the limit of $d\rightarrow0$, $F_1(d)\sim d^2 \rightarrow0$, in stark contrast with the case of perfect detection. Therefore, in realistic situation with noisy detections, the super-resolution feature of binary SPADE measurement with arbitrarily close separation is lost. For illustration, the formulae Eq.~(\ref{eq22}) and Eq.~(\ref{eq23}) with different $\SNR$ are plotted in Fig.~\ref{Fig4}.  \medskip

While $F_1(d)\rightarrow0$ for $d\rightarrow0$, fortunately, however, we may still have near optimally precise estimators, for a range of small but \emph{finite} separation. On one hand, consistent with our interest in the small separation regime, we have $d/\sigma\ll1$, such that Eq.~(\ref{eq22}) and Eq.~(\ref{eq23}) are valid in the first place. On the other hand, there might exist a finite range of $d$ such that the factor $\big(1+\frac{4\sigma^2}{d^2}\beta\big)^{-1}$ is close to one, i.e., $\frac{4\sigma^2}{d^2}\beta\ll1$ or
\begin{align}
d\gg \frac{2\sigma}{\sqrt{\SNR}}. \label{eq24}
\end{align}
Combining the two inequalities, we thus have
\begin{align}
\frac{2\sigma}{\sqrt{\SNR}}\ll d \ll \sigma\label{eq25}
\end{align}
as the range of $d$ for which the performance of binary SPADE measurement is not affected much by the noise for Poissonian sources, such that $F_1^{\textsc{(p)}}(d)\approx\FQd=\frac{\ns}{\sigma^2}$. \medskip

For thermal sources, as in the noiseless case, we need the additional condition $d/\sigma\ll2/\sqrt{\ns}$ such that $F_1^{\textsc{(t)}}(d)\approx\FQd$. More generally, when the detected mean photon number is low, in the sense of $\ns\tau_1\ll1$ and $\nb\ll1$, such that $\kbar=\ns\tau_1+\nb\ll1$, we can as before consider only the single-photon events which happen with probability $p_1\approx\kbar$, and regardless of the photon statistics, obtain the FI
\begin{align}
F_1(d)&\approx \frac{1}{p_1}\Big(\frac{\partial}{\partial d}p_1\Big)^2\approx \frac{\ns}{\sigma^2}\Big( \frac{1}{1+\frac{4\sigma^2}{d^2}\beta} \Big).&&[\ns\tau_1\ll1, \nb\ll1]\label{eq26}
\end{align}
The range of $d$  where we still have superresolution with binary SPADE measurement is then given by, after combining all the conditions,
\begin{align}
\frac{2\sigma}{\sqrt{\SNR}}\ll d\ll \min\Big(\sigma,\frac{2\sigma}{\sqrt{\ns}}\Big). && [\ns\tau_1\ll1, \nb\ll1] \label{eq27}
\end{align}
Of course, Eq.~(\ref{eq25}) and Eq.~(\ref{eq27}) are only meaningful if the inequalities hold consistently, i.e., $\SNR\gg 1$ \emph{a priori}.

The upper bounds in Eq.~(\ref{eq25}) and Eq.~(\ref{eq27}) are essential the same condition that is present also for the noiseless binary SPADE. It specifies the valid range of $d$ before we need to consider the measurements with higher order modes. Let us then focus on the new lower bound that is not present in the perfect case, which specifies the smallest $d$ which we can estimate for realistic binary SPADE measurement with near-optimal precision. To provide a benchmark to the resolution limit, we introduce the \emph{half-resolution distance} $\dhalf$ as the minimum separation for which the FI drops to about half of its maximum value, i.e.,
\begin{align}
\dhalf\equiv\frac{2\sigma}{\sqrt{\SNR}}; \quad \quad  F_1(\dhalf)\approx\frac{\FQd}{2}=\frac{\ns}{2\sigma^2}. && [\nb\ll1] \label{eq28}
\end{align}
Accordingly, if our resources are restricted such that we can achieve at most a certain $\SNR$ in the experiment, $\dhalf$ sets the limit of separation for which we could estimate with a precision that is \emph{at least} a fraction of $1/\sqrt{2}$ of the perfect case; smaller $d$ would have worse precision. An alternate view point is, if we would like to achieve superresolution for some separation $d$ (which is $\ll\sigma$), we must then ensure that we have a large $\SNR$, one which is much greater than $4\sigma^2/d^2$. Moreover, it is important to note that, not only $\SNR\gg4\sigma^2/d^2$ is needed, in general one must have $\nb\ll1$ in $T$ as well. If $\nb$ is sufficiently large, then increasing the $\SNR$ might not improve the resolution beyond a certain fraction of $\FQd$, we have seen for the case with thermal sources.

\section{Binary SPADE with quadrature measurements}\label{Sec5}
Other than photon-counting measurement, popular choices of measurement in optics experiments include the field quadrature measurements. In the context of resolving two light sources, the performance of binary SPADE with quadrature measurements has been studied for example in Refs.~\citen{YangTashilinaOPTICA2016,YangNair2017}.  In spite of the ever present shot noise, Ref.~\citen{YangNair2017} demonstrated that for thermal sources, binary SPADE with quadrature measurements still offer estimation with finite FI over some range of $d$, and outperforms direct imaging when the signal is strong enough. In this section, we will provide a simple quantification of the noise in the case with homodyne and heterodyne measurements, and then similarly introduce the characteristic separation $\dhalf$ that benchmark the resolution limits. \medskip

The filtering of the light into the mode $v_1(x)$ amounts to a reduction of mean photon number from $\ns$ to $\ns\tau_1(d)$. We will characterize quadratures in units such that shot noise generates variance $1/2$. In the case of the homodyne detection of a single quadrature, the thermal signal will contribute excess variance $\ns\tau_1(d)$. In the case of heterodyne (double homodyne) detection, each quadrature will exhibit variance $1/2+\ns\tau_1(d)/2$, as the input signal power is now split equally between both measured quadratures.

\subsection{Homodyne measurement}
Let us first consider the homodyne measurement, where we measure just one quadrature. The probability of obtaining the quadrature value $q$ is then \cite{Leonhardt1997}
\begin{align}
p(q)&=\frac{1}{\sqrt{\pi}\sqrt{1+2\ns\tau_1}}\upe^{-\frac{q^2}{1+2\ns\tau_1}}. \label{eq29}
\end{align}
As $p(q)$ is Gaussian, we can make use of the results Eq.~(\ref{eqA5}) and Eq.~(\ref{eqA6}) in Appendix, and obtain the FI
\begin{align}
F_1^\textsc{(hom)}(d)\approx \frac{2\ns^2\Big(\frac{\partial}{\partial d}\tau_1\Big)^2}{\big(1+2\ns\tau_1\big)^2}\approx \frac{2\ns^2 d^2}{\big( \ns d^2+2\sigma^2 \big)^2}\label{eq30}.
\end{align}
In the limit of $d\rightarrow0$, we have $F_1^\textsc{(hom)}(d)\sim \frac{\ns^2 d^2}{2\sigma^4}\rightarrow0$, and as has been discussed also in Ref.~\citen{YangNair2017}, the maximum of FI in Eq.~(\ref{eq30}) is only 1/4 of $\FQd$.

The fact that $F_1^\textsc{(hom)}(d)\rightarrow0$ can be attributed to the presence of shot noise, where unlike the (perfect) photon-counting measurement, the statistics of the quadrature $q$ given in Eq.~(\ref{eq29}) always exhibits non-zero variance, which in the absence of the signal tends to the value 1/2 which we will take as the noise figure.
Identifying $\ns$ as the signal, our $\SNR$ is thus $2\ns$. As the maximum FI for homodyne detection is $\FQd/4$, we define the characteristic length as the smallest separation, such that $F_1^\textsc{(hom)}(\dhalf)\approx \FQd/8$, i.e.,
\begin{align}
\dhalf\equiv \frac{2-\sqrt{2}}{\sqrt{\ns}}\sigma=\frac{2\sqrt{2}-2}{\sqrt{\SNR}}\sigma. \label{eq31}
\end{align}
For illustration, in Fig.~\ref{Fig5}, we plot the graphs of exact FI per mean incoming source photon with homodyne measurement with different $\SNR$, and $\dhalf$ that approximates the separation such that $F^\textsc{(hom)}_1(d)\approx \FQd/8$.
\begin{figure}[t!]
	\centering
	\includegraphics[width=0.9\columnwidth]{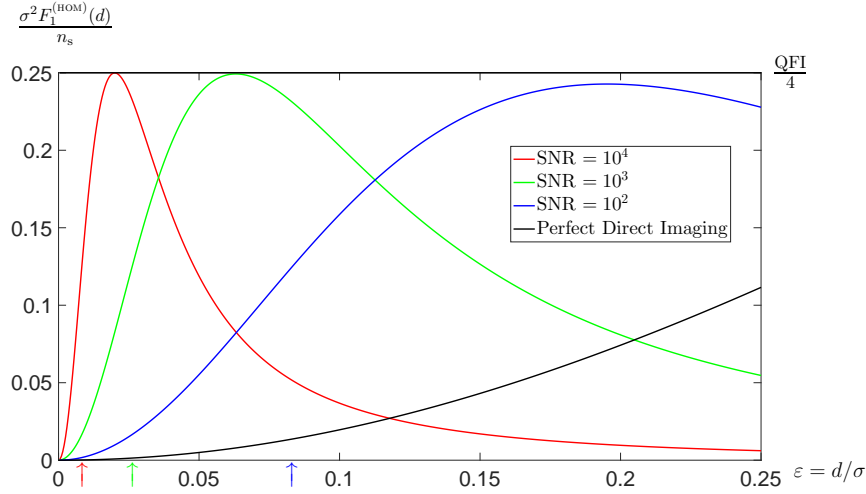}
	\caption{Binary SPADE with homodyne detection: FI per mean incoming source photon $\ns$, multiplied by $\sigma^2$, for different signal-to-noise ratios (SNR), with Gaussian transfer function. The presence of shot noise causes the loss of superresolution for arbitrarily small $d$, and we characterize the resolution limits by $\dhalf\equiv\frac{(2\sqrt{2}-2)\sigma}{\sqrt{\SNR}}$, which is the smallest separation that can be estimated with at least $\frac{1}{\sqrt{2}}$ of the optimal precision, which in this case, corresponds to only $1/4$ of the QFI. In the figure, they are marked by the three arrows with the corresponding color codes. For comparison, we have included the FI computed for perfect direct imaging (the black solid curve).}
	\label{Fig5}
\end{figure}

\subsection{Heterodyne measurement}
For heterodyne or double homodyne measurement, we split the light into two by a 50:50 beam splitter, and then measure simultaneously two conjugate quadratures. The probability density of obtaining values $q$ and $p$ for the two quadrature measurements is then
\begin{align}
p(q,p)&=\frac{1}{\pi(1+\ns\tau_1)}\upe^{-\frac{q^2+p^2}{1+\ns\tau_1}}. \label{eq32}
\end{align}
As $p(q,p)$ is a two-variate Gaussian, using Eq.~(\ref{eqA7}) and Eq.~(\ref{eqA8}) in Appendix, we obtain the FI
\begin{align}
F_1^\textsc{(het)}(d)\approx \frac{\ns^2\Big(\frac{\partial}{\partial d}\tau_1\Big)^2}{\big(1+\ns\tau_1\big)^2}\approx \frac{4\ns^2 d^2}{\big( \ns d^2+4\sigma^2 \big)^2}\label{eq33}.
\end{align}
Comparing Eq.~(\ref{eq33}) to Eq.~(\ref{eq30}), there is no qualitative difference between homodyne and heterodyne measurement. The largest obtainable FI is still $\FQd/4$. The noise figure is obtained by combining the variances of both quadratures measured in the absence of the signal and equals 
\begin{align}
\int_{-\infty}^\infty \mathrm{d}q\, \mathrm{d}p\, \frac{\upe^{-q^2-p^2}}{\pi} (q^2+p^2)=1, \label{eq34}
\end{align}
which is twice of that for homodyne detection. Thus, we have $\SNR=\ns$, and the smallest separation such that $F_1^\textsc{(het)}(\dhalf)\approx \FQd/8$ is
\begin{align}
\dhalf\equiv \frac{2\sqrt{2}-2}{\sqrt{\ns}}\sigma=\frac{2\sqrt{2}-2}{\sqrt{\SNR}}\sigma, \label{eq35}
\end{align}
which is exactly the same expression as in Eq.~(\ref{eq31}).

\section{Conclusion}\label{Sec6}
In summary, in this work, we have studied the resolution limits of binary SPADE, in the presence of noise. We found that the super-resolution feature for arbitrarily small separation is lost, and as a rule of thumb, we introduce the minimum resolvable spatial separation $\dhalf$ which achieves half of the maximum FI for different choices of measurement strategy. For both the photon-counting and quadrature measurements, we show that $\dhalf$ has a characteristic dependence of $\sim\SNR^{-1/2}$.

\section*{Acknowledgments}
We acknowledge insightful discussions with M.~Jarzyna, J.~Ko{\l}ody\'{n}ski, A.~Lvovsky, and N.~Treps. M.~P. was supported by the Foundation for Polish Science (FNP). This work is part of the project ``Quantum Optical Communication Systems'' carried out within the TEAM programme of the Foundation for Polish Science co-financed by the European Union under the European Regional Development Fund. Y.~L. L. and C.~D. were supported by the project ``Quantum Optical Technologies"€ carried out within the International Research Agendas programme of the Foundation for Polish Science co-financed by the European Union under the European Regional Development Fund.

\appendix
\section{Fisher Information}\label{Appen}
In this appendix, we supplement the details of our calculations for the FI in Eq.(\ref{eq10}) and Eq.(\ref{eq11}), as well as Eq.~(\ref{eq30}) and Eq.~(\ref{eq33}).

\subsection{\textbf{Poissonian photocount statistics}}
Suppose the photon number statistics of light arriving from the two mutually incoherent sources is Poissonian with the mean number $\ns$ over the detection time. Under the assumption that the detection acts are statistically independent, the probability of detecting $k$ photons in the mode $v_1(x)$ is
\begin{align}
p_k=\sum_{j=0}^\infty \upe^{-\ns}\frac{\ns^j}{j!} \binom{j}{k} \tau_1^k (1-\tau_1)^{j-k}\equiv \upe^{-\kbar}\frac{(\kbar)^k}{k!}, \label{eqA1}
\end{align}
which is exactly a Poissonian with mean photon number $\kbar=\ns\tau_1(d)$. Generally, the FI for a Poisson probability distribution with mean $\kbar$ is
\begin{align}
F^{\textsc{(p)}}(d)=\sum_{k=0}^\infty \frac{1}{p_k}\Big(\frac{\partial}{\partial d} p_k\Big)^2=\frac{1}{\kbar}\Big(\frac{\partial}{\partial d}\kbar\Big)^2,  \label{eqA2}
\end{align}
which holds true for any functional dependence $\kbar(d)$.

\subsection{\textbf{Bose-Einstein photocount statistics}}
Another relevant scenario is photodetection of a single temporal mode of light generated by sources in thermal equilibrium. With the average photon number $\ns$, the photocount statistics for the mode $v_1(x)$ is
\begin{align}
p_k=\sum_{j=0}^\infty \frac{1}{\ns+1}\Big(\frac{\ns}{\ns+1}\Big)^{j} \binom{j}{k} \tau_1^k (1-\tau_1)^{j-k}\equiv \frac{1}{\kbar+1}\Big(\frac{\kbar}{\kbar+1}\Big)^{k}, \label{eqA3}
\end{align}
which is again a Bose-Einstein distribution, with now mean photon number $\kbar=\ns\tau_1(d)$. Take note that, while in the case of Poissonian and Bose-Einstein photocount statistics non-unit transmission reduces their mean but preserves their characteristics, this is not generally true for other probability distributions.

\noindent The FI for a Bose-Einstein probability distribution with mean $\kbar$ is
\begin{align}
F^{\textsc{(t)}}(d)=\sum_{k=0}^\infty \frac{1}{p_k}\Big(\frac{\partial}{\partial d} p_k\Big)^2=\frac{1}{\kbar+1}\frac{1}{\kbar}\Big(\frac{\partial}{\partial d}\kbar\Big)^2 = \frac{1}{\kbar+1} F^{\textsc{(p)}}(d),  \label{eqA4}
\end{align}
which, as compared to the Poissonian case, has an additional factor of $1/(\kbar+1)$. As this factor is always smaller than one, $F^{\textsc{(t)}}(d)$ is always smaller than $F^{\textsc{(p)}}(d)$ for the same mean $\kbar$.

\subsection{\textbf{Quadrature measurements}}
In addition to photon-counting measurement, one can also perform measurement of the electromagnetic field quadratures by means of phase-sensitive detection. For homodyne detection which measures a single quadrature with a Gaussian probability density function with variance $V(d)$,
\begin{align}
p(x)&=\frac{1}{\sqrt{2\pi V(d)}} \upe^{-\frac{x^2}{2V(d)}}, \label{eqA5}
\end{align}
the FI is
\begin{align}
F^\textsc{(hom)}(d)&=\int_{-\infty}^{\infty}\,\mathrm{d}x\, \frac{1}{p(x)}\Big(\frac{\partial}{\partial d}p(x)\Big)^2=\frac{1}{2V^2}\Big(\frac{\partial}{\partial d}V\Big)^2. \label{eqA6}
\end{align}
In particular, we have $V=1/2+\ns\tau_1$ for Eq.~(\ref{eq29}) and Eq.~(\ref{eq30}). \medskip

\noindent For heterodyne detection which measures both the quadratures, with the joint Gaussian probability density
\begin{align}
p(x,y)&=\frac{1}{2\pi V(d)} \upe^{-\frac{x^2+y^2}{2V(d)}}, \label{eqA7}
\end{align}
the FI is
\begin{align}
F^\textsc{(het)}(d)&=\int_{-\infty}^{\infty}\,\mathrm{d}x\, \int_{-\infty}^{\infty}\,\mathrm{d}y\, \frac{1}{p(x,y)}\Big(\frac{\partial}{\partial d}p(x,y)\Big)^2=\frac{1}{V^2}\Big(\frac{\partial}{\partial d}V\Big)^2 = 2F^\textsc{(hom)}(d). \label{eqA8}
\end{align}
In particular, we have $V=1/2+\ns\tau_1/2$ for Eq.~(\ref{eq32}) and Eq.~(\ref{eq33}). Note that if $V(d)$ in Eqs.~(\ref{eqA5}) and (\ref{eqA7}) are the same one has $F^\textsc{(het)}(d) = 2F^\textsc{(hom)}(d)$ which follows immediately from the additivity of Fisher information.

\bibliographystyle{ws-ijqi}

\end{document}